%% file: 00-main.tex
\documentclass[lettersize,journal]{IEEEtran}

\usepackage{amsmath,amsfonts}
\usepackage{algorithmic}
\usepackage{algorithm}
\usepackage{array}
\usepackage[caption=false,font=normalsize,labelfont=sf,textfont=sf]{subfig}
\usepackage{textcomp}
\usepackage{stfloats}
\usepackage{url}
\usepackage{verbatim}
\usepackage{graphicx}
\usepackage{cite}
\usepackage{bm}
\usepackage[table]{xcolor}
\usepackage{multirow}
\usepackage{tikz}
\usepackage{moresize}

\usepackage[bookmarks=true,breaklinks=true,letterpaper=true,colorlinks,citecolor=blue,linkcolor=blue,urlcolor=blue]{hyperref}

\newcommand{\tinytilde}{\raisebox{0.5ex}{\scalebox{0.7}{$\sim$}}}

\DeclareRobustCommand{\whitecircle}[1]{%
  \tikz[baseline=(char.base)]{
    \node[shape=circle, draw=black, fill=white,
          inner sep=1pt, line width=0.5pt] (char) {#1};
  }%
}

\definecolor{myblue}{RGB}{106,154,208}

\DeclareRobustCommand{\bluecircle}[1]{%
  \tikz[baseline=(char.base)]{
    \node[shape=circle, draw=black, fill=myblue,
          inner sep=1.2pt, line width=0.5pt] (char) {#1};
  }%
}

\begin{document}
% -------------------------------------------------------------------------------

\title{Performance and Energy Benefits of MRDIMMs}

\author{Pau Díaz, Mariana Carmin, Pouya Esmaili-Dokht, Victor Xirau, Felippe Zacarias, Henrique Potter, Harald Servat, Miquel Moreto, Eduard Ayguadé, Petar Radojkovi\'{c}

\thanks{P. Díaz, V. Xirau and P. Radojkovic are with Barcelona Supercomputing Center (e-mail: pau.diazcuesta@bsc.es; vxirau@bsc.es; petar.radojkovic@bsc.es).}
\thanks{M. Carmin, P. Esmaili-Dokht, M. Moreto and E. Ayguadé are with Barcelona Supercomputing Center and Universitat Politècnica de Catalunya (e-mail: name.surname@bsc.es).}
\thanks{F. Zacarias and H. Potter are with Micron Technology (e-mail: fvieirazacar@micron.com; hpotter@micron.com).}
\thanks{H. Servat is with Intel Corporation (e-mail: harald.servat@intel.com).}
\vspace{-2.0ex}}

\maketitle

% -------------------------------------------------------------------------------
\begin{abstract}
Multiplexed Rank DIMMs (MRDIMMs) have recently emerged as memory devices that enable higher bandwidth without increasing DRAM chip frequencies. 
This paper presents a detailed performance, power and energy evaluation of a production server with high-end MRDIMM main memory. 
The memory system upgrade from conventional registered DIMMs~(RDIMMs) to MRDIMMs extends the bandwidth by 41\% 
yielding 27--41\% higher performance for bandwidth-bound workloads. 
Additionally, the latency improvement reaches hundreds of nanoseconds, benefiting a broad class of workloads sensitive to memory latency. 
At the same bandwidth utilization levels, RDIMMs and MRDIMMs exhibit similar power consumption.   
In the MRDIMM-extended bandwidth region, the performance improvements largely exceed the power increase,  
delivering up to 30\% server energy savings for memory-bound workloads.

\end{abstract}

% -------------------------------------------------------------------------------

% -------------------------------------------------------------------------------
\input{10-Introduction}
% -------------------------------------------------------------------------------

% -------------------------------------------------------------------------------
% Section: MRDIMM design innovations
\input{20-Background}

% -------------------------------------------------------------------------------

% -------------------------------------------------------------------------------
% Section: Experimantal 
\input{25-Experimental}
% -------------------------------------------------------------------------------

% -------------------------------------------------------------------------------
% Section: MRDIMMs: Outperforming RDIMMs in the whole memory bandwdith range
\input{30-Performance}

% -------------------------------------------------------------------------------

% -------------------------------------------------------------------------------
% Section: In power the price to pay? 
\input{40-Power}

% -------------------------------------------------------------------------------

% -------------------------------------------------------------------------------
% Section: Energy efficiency 
\input{50-Energy}

% -------------------------------------------------------------------------------

% -------------------------------------------------------------------------------
% Section: Adoption
\input{60-MRDIMM-adoption}

% -------------------------------------------------------------------------------

% -------------------------------------------------------------------------------
% Section: Conclusions
\input{70-Conclusions}
% -------------------------------------------------------------------------------

% -------------------------------------------------------------------------------
\bibliographystyle{unsrt}
\bibliography{refs}
% -------------------------------------------------------------------------------

\end{document}

%% file: 10-Introduction.tex
% -------------------------------------------------------------------------------
\section{Introduction}
\label{sec:introduction}
% -------------------------------------------------------------------------------

\IEEEPARstart{I}{n} conventional DIMMs, the DRAM chips and the host interface operate at the same frequency. 
Multiplexed Rank DIMMs~(MRDIMMs) perform multiplexing between the host memory channel and the DRAM chips,  
allowing the DRAM chips to operate at their native data rate while \textbf{doubling} the memory-channel frequency. 
This simple design innovation
enables significant improvements in \textbf{memory performance, energy efficiency, and capacity}---benefits that previously required multiple generations of DRAM technology scaling~\cite{Intel:MRDIMM-news}.

\looseness -1 This paper investigates the implications of upgrading a production server from a conventional DDR5 RDIMM--6400 to high-end MRDIMM--8800 memory.   
Gen.\,2 MRDIMMs, planned for 2026/2027, are expected to reach 12,800\,MT/s\,\cite{Lenovo:MRDIMMs}. 
 
The MRDIMMs show two key performance advantages (Sec.\,\ref{sec:performance}). First, they unlock an \textbf{extended memory-bandwidth window}.
The sustained memory bandwidth increases by 41.3\%, from around 500\,GB/s (12$\times$RDIMM--6400)
to more than 700\,GB/s (12$\times$MRDIMM--8800).  %, an increment of 41.3\%. 
This is fully exploited by bandwidth-bound workloads, which achieve 27--41\% higher performance. 
Second, the MRDIMMs \textbf{reduce memory access latency} by up to tens of percent, or \textbf{hundreds of nanoseconds}. 
In the platform under study, the latency improvement is visible in the entire range of memory-bandwidth utilization, from the unloaded to the fully-saturated memory systems, therefore benefiting a broad class of latency-sensitive workloads.

Early studies report substantial increases in power consumption in servers equipped with MRDIMMs~\cite{phoronix:MRDIMMs, Dravai:MRDIMMs}. This contributed to the prevailing perception that the performance gains enabled by
MRDIMMs come at a significant power cost.  
Our power analysis in Sec.\,\ref{sec:power} distinguishes between two bandwidth regions: the shared window,  
in which both RDIMM- and MRDIMM-based systems operate, and
the MRDIMM-extended window,  
which is enabled by the RDIMM-to-MRDIMM upgrade.
Across most of the shared bandwidth window, \textbf{MRDIMMs exhibit similar or even lower power
consumption} than RDIMMs at the same level of memory-bandwidth utilization. 
In the performance window unlocked by the MRDIMM upgrade, the system does consumes more power. 
However, this power increase is driven by higher delivered performance, 
rather than by intrinsically power-hungry MRDIMMs, as is often assumed~\cite{phoronix:MRDIMMs}.   
Actually, in the MRDIMM-extended bandwidth window, the performance improvement largely exceeds the power increase, 
making it \textbf{the most energy-efficient operating region}. 
The benchmarks that exploit this region reduce their energy consumption by 19--30\%~(Sec.\,\ref{sec:energy}).  

%% file: 20-Background.tex
% -------------------------------------------------------------------------------
\section{MRDIMM design innovations}
\label{sec:background}
% -------------------------------------------------------------------------------

% ---------------------------------
% MRDIMM tech innovation 
% ---------------------------------
\begin{figure}[b!]
    \vspace{-2ex}
    \centering
    \subfloat[{RDIMMs and LRDIMMs: A registering clock driver (RCD)~\whitecircle{1} and, in the case of LRDIMMs, data buffers (DBs)~\whitecircle{2} buffer and regenerate signals between the memory channel and the DRAM chip. The DRAM chips and the host interface operate at the same frequency~\whitecircle{3}.}\vspace{-1ex}\label{fig:R-LRDIMM-arch} ]
    {\includegraphics[width=1\columnwidth]{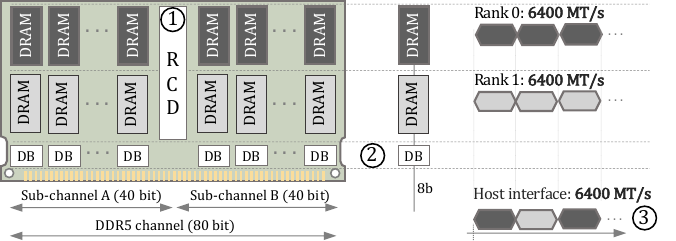}}

    \subfloat[{MRDIMMs: A multiplexing RCD~(MRCD)~\bluecircle{1} and multiplexing data buffers~(MDBs)~\bluecircle{2} time-multiplex command, address and data signals between the memory channel and the DRAM chips. MRDIMMs communicate with the host at \textbf{twice} the native DRAM frequency~\bluecircle{3}.}\label{fig:MRDIMM-arch}]
    {\includegraphics[width=1\columnwidth]{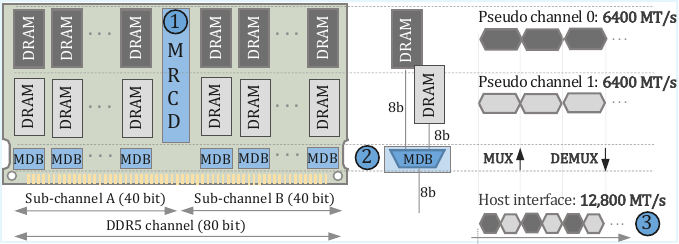}}
    
    \caption{Architectural comparison of RDIMMs/LRDIMMs and MRDIMMs: Simple MRDIMM design enhancements enable a significant increase in memory bandwidth at the memory-controller interface.} 
    \label{fig:rdimms-vs-mrdimms-arch}
\end{figure}

The capacity and performance of memory devices are fundamentally constrained by signal integrity~\cite{rambus:mrdimms}. \textbf{Registered DIMMs~(RDIMMs)}, illustrated in Fig.\,\ref{fig:R-LRDIMM-arch}, improve signal quality by introducing a registering clock driver~(RCD)\,\whitecircle{1} that buffers and regenerates command and address signals from the memory channel and distributes them to the DRAM chips\,\cite{JEDEC:JESD82-514_2024}. \textbf{Load-Reduced DIMMs\,(LRDIMMs)} extend this approach by placing data buffers~(DBs)\,\whitecircle{2} on the data path between the host interface and the DRAM devices\,\cite{JEDEC:JESD82-521_2021}. 
In RDIMM- and LRDIMM-based memory systems, the host interface operates at the same frequency as the DRAM devices\,\whitecircle{3}\, and the host can access only a single DRAM rank at a time~\cite{rambus:mrdimms}. 

In \textbf{Multiplexed Rank DIMMs~(MRDIMMs)}, shown in Fig.\,\ref{fig:MRDIMM-arch}, 
each sub-channel is divided into two independent pseudo-channels with separate commands, addresses and data paths.  
Signals from the independent pseudo-channels are time-multiplexed onto the same bus from the host memory controller by an enhanced registering clock driver and multiplexing data buffers~\cite{rambus:mrdimms}. 
The multiplexing RCD~(MRCD)~\bluecircle{1} demultiplexes signals received from the memory controller interface, operating, for example, at 12,800\,MT/s, and concurrently delivers independent command and address streams to each MRDIMM pseudo-channel running at 6400\,MT/s. 
On the data path, multiplexing data buffers~(MDBs) perform multiplexing and demultiplexing between the memory channel and the DRAM chips~\bluecircle{2}. Overall, MRDIMM multiplexing allows the DRAM chips to operate at their native data rate while \textbf{doubling} the effective memory-channel frequency~\bluecircle{3}.

MRDIMM innovations also enable significantly higher capacities. 
In current RDIMM designs, scaling beyond two ranks per DIMM is challenging 
because it increases electrical loading on the memory controller and introduces signal-integrity challenges on the CPU-to-memory bus.  
MRDIMMs overcome this limitation by deploying two ranks per pseudo-channel (not illustrated in Fig.\,\ref{fig:MRDIMM-arch}), providing a simple and cost-efficient way to double the number of ranks and the overall DIMM capacity~\cite{rambus:mrdimms, Micron:MRDIMMs}. 
The higher memory capacity, together with a higher per-node throughput
often translates into server consolidation and a lower total cost of ownership, which is discussed in Sec.\,\ref{sec:MRDIMM-adoption}.   

One of MRDIMM’s key advantages is that it can serve as a \textbf{drop-in replacement} for server memory upgrades. The server evaluated in this work, as well as other forthcoming DDR5 CPUs and platforms, supports both RDIMM and MRDIMM. As a result, users are not required to choose between RDIMM and MRDIMM at the initial design and deployment stage. This flexibility extends across the server lifecycle, enabling data centers to adopt MRDIMM during later upgrade cycles when higher memory performance is needed~\cite{rambus:mrdimms}.

%% file: 25-Experimental.tex
% -------------------------------------------------------------------------------
\section{Experimental environment}
\label{sec:experimental}
% -------------------------------------------------------------------------------

\looseness -1
We study the implications of upgrading 
a dual-socket Intel\textsuperscript{\textregistered} Xeon\textsuperscript{\textregistered} 6980P (Granite Rapids) server~\cite{xeon-6980p, xeon_6} from DDR5 RDIMM--6400 to MRDIMM--8800 main memory.  

Each CPU comprises 128\,cores operating in Latency Optimized mode, with a maximum frequency of 3.2\,GHz. 
The CPU has 12~DDR5 memory channels populated with one dual-rank DIMM per channel in all experiments. 
We quantify the benefits of Gen.\,1 MRDIMM devices operating at 8800\,MT/s. % 
To ensure a fair comparison of performance, power and energy, we evaluate RDIMMs and MRDIMMs with identical 
capacities of 64\,GB. Throughout the paper, we report performance, power and energy measurements on a per-socket basis.  

Performance, power and energy consumption of the RDIMM- and MRDIMM-based systems are evaluated with 
an extensive set of memory-bound benchmarks. 
Performance measurements are based on the metrics defined by each of the benchmarks: 
sustained memory bandwidth for STREAM~\cite{mccalpin:streamBenchmark}, Google Multiload~\cite{Google:multichase}, and LMbench~\cite{lmbench}; floating-point operations per second for HPCG~\cite{intel:hpcg}; unloaded memory latency for Intel MLC~\cite{Intel:MLC} and Google Multichase~\cite{Google:multichase}; and bandwidth--latency curves for 
a recently-released Mess benchmark~\cite{esmailidokht2024mess}. 
The Mess benchmark generates complex memory-traffic patterns determined by sequential accesses within each core as well as by interleaving of memory requests across cores~\cite{esmailidokht2024mess}. To validate our findings across different access patterns, we repeated all experiments using the Mess-Random benchmark, which generates random traffic at each core. 
All the paper’s conclusions remained unchanged. 

DRAM and CPU power and energy are measured using RAPL~\cite{rapl} counters. Total server power consumption is measured using IPMItool~\cite{ipmitool}, which reads hardware sensor data through the Baseboard Management Controller.

%% file: 30-Performance.tex
% -------------------------------------------------------------------------------
\section{MRDIMMs: Outperforming RDIMMs in the whole memory bandwidth range}
\label{sec:performance}
% -------------------------------------------------------------------------------

\begin{figure}[t!]
    \centering
    
    \includegraphics[width=1\columnwidth]{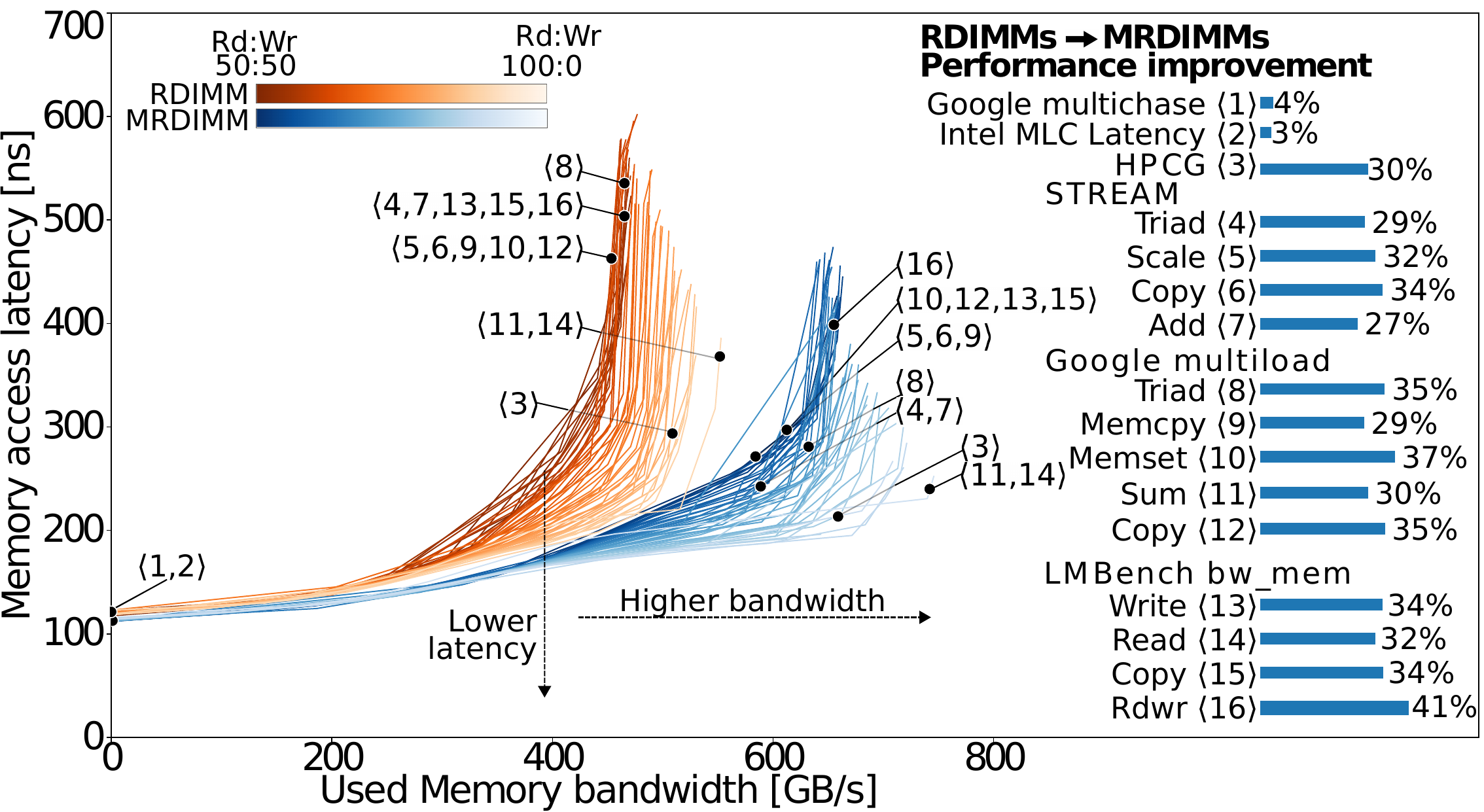}
    \caption{Comparison of an Intel\textsuperscript{\textregistered} Xeon\textsuperscript{\textregistered} 6980P CPU with 12$\times$RDIMM--6400 \textit{vs.} 12$\times$MRDIMM--8800 devices. The higher theoretical MRDIMM bandwidth is reflected in the measured bandwidth--latency curves and in the performance gains of bandwidth-bound workloads. Also, MRDIMMs exhibit lower memory-access latency across the entire bandwidth-utilization range.}
    \label{fig:mrdimms-vs-rdimms:performance}

    \vspace{3ex}

    \begin{minipage}[t]{0.54\linewidth}
        \vspace{0pt}
        \includegraphics[width=\linewidth]{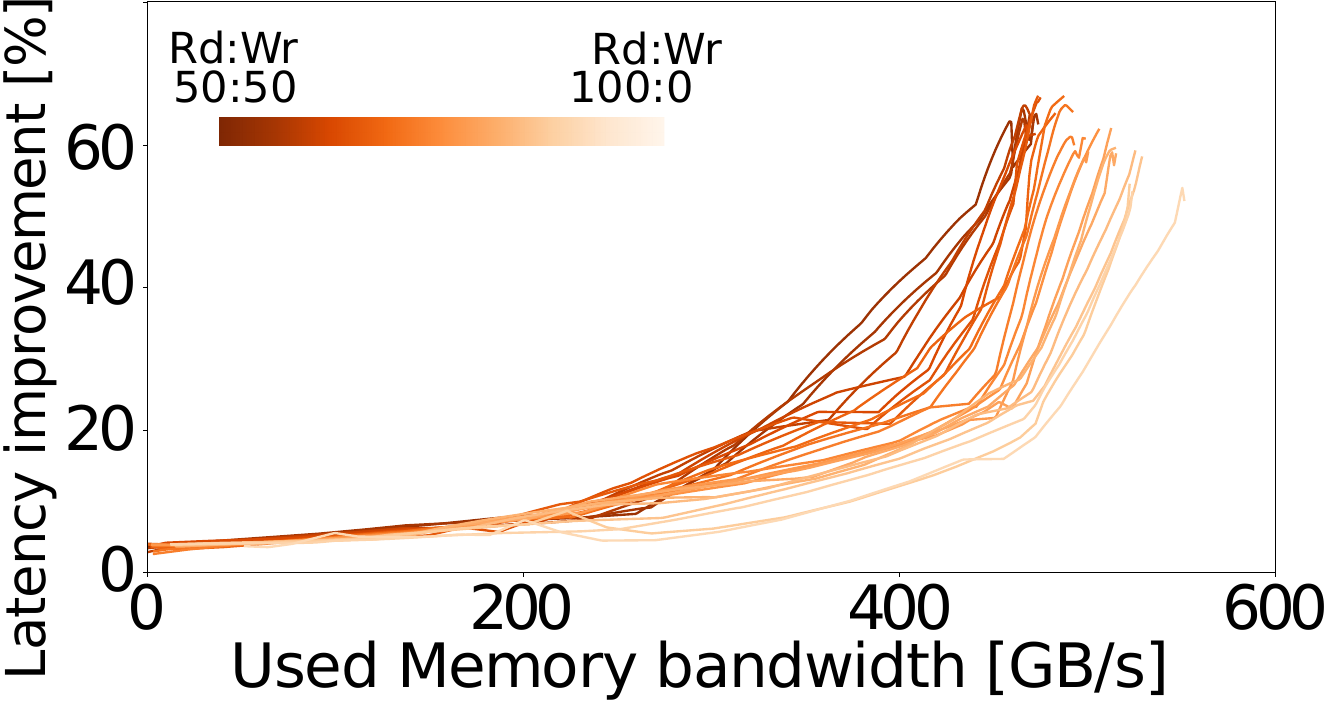}
    \end{minipage}\hfill
    \begin{minipage}[t]{0.44\linewidth}
        \captionsetup{singlelinecheck=on, skip=0pt, justification=raggedright}
        \caption{Latency improvement of the RDIMM-to-MRDIMM upgrade increases with the bandwidth utilization. It reaches tens of percent, or hundreds of nanoseconds, in the saturated memory region.}
        \label{fig:latency_change}
    \end{minipage}
\end{figure}

Fig.\,\ref{fig:mrdimms-vs-rdimms:performance} compares the RDIMM and MRDIMM memory system performance.  
The bandwidth–latency curves, measured with the Mess benchmark, 
plot memory access latency ($y$-axis) from an unloaded to a fully saturated memory system ($x$-axis). Multiple compositions of read and write traffic are plotted with different shades of orange~(RDIMMs) and blue~(MRDIMMs).  

The horizontal bar chart in Fig.\,\ref{fig:mrdimms-vs-rdimms:performance} also shows the performance improvements of the memory-bound benchmarks. 
We use angle brackets, e.g. $\langle$3$\rangle$ for HPCG,  
to show the position of the benchmarks in the memory bandwidth--latency curves. 
Fig.\,\ref{fig:mrdimms-vs-rdimms:performance}, therefore, connects the memory system performance (bandwidth--latency curves), the benchmarks' memory system utilization (position in the curves), and the performance improvements (bar chart). 

Increasing the memory channel data rate from RDIMMs--6400 to MRDIMM--8800 raises theoretical memory bandwidth by 37.5\%, from 51.2\,GB/s to 70.4\,GB/s per channel. Sustained bandwidth, measured by the Mess benchmark, improves by up to 41.3\%, slightly exceeding the theoretical increase because of better memory-bus utilization.    
The bandwidth-bound benchmarks, $\langle$3--16$\rangle$ in Fig.\,\ref{fig:mrdimms-vs-rdimms:performance}, 
successfully exploit the higher sustained MRDIMM bandwidth and achieve 27--41\% performance improvements.  

In the platform under study, the MRDIMM-based memory system shows 
lower latency across the entire bandwidth range over which both RDIMM- and MRDIMM-based systems operate. 
The latency benefit of the RDIMM-to-MRDIMM upgrade, shown in Fig.\,\ref{fig:latency_change}, increases with the bandwidth utilization because of lower queuing overheads, reaching tens of percent, or hundreds of nanoseconds, in the saturated memory region. 

The latency-sensitive Intel MLC and Google Multichase benchmarks, $\langle$1,\,2$\rangle$ in Fig.\,\ref{fig:mrdimms-vs-rdimms:performance}, show 3--4\% performance improvements, 
corresponding to an approximately 5\,ns reduction in unloaded memory access latency in the MRDIMM-based system. This result is counterintuitive, as MRDIMMs incorporate multiplexing data buffers (Fig.\,\ref{fig:MRDIMM-arch})\textbf{} that are expected to introduce additional latency~\cite{rambus:mrdimms}.

Discussions with CPU and memory vendors suggest that this behavior is 
is not due to the intrinsic MRDIMM access latency, but rather to differences in how the processor schedules and services memory requests in RDIMM- and MRDIMM-based systems. In particular, the higher memory-channel frequency and the presence of independent pseudo-channels in MRDIMMs reduce queueing conflicts and increase effective parallelism in the memory subsystem. These effects enable the processor to leverage more effective latency-optimization mechanisms, resulting in improved performance for latency-sensitive workloads across the entire bandwidth range.

%% file: 40-Power.tex
% -------------------------------------------------------------------------------
\section{Is power the price to pay?}
\label{sec:power}
% -------------------------------------------------------------------------------
Early studies report substantial increases in power consumption in servers equipped with  MRDIMMs. 
Drav\'ai and Reguly~\cite{Dravai:MRDIMMs} evaluate memory-bound HPC benchmarks and observe approximately a 50\% increase in both DRAM and total server power when moving from Intel\textsuperscript{\textregistered} Xeon\textsuperscript{\textregistered} Platinum 8592+ (Emerald Rapids) server with DDR5--5600 RDIMMs to 
Intel\textsuperscript{\textregistered} Xeon\textsuperscript{\textregistered} Platinum 6960P (Granite Rapids) server with DDR5--8800 MRDIMMs.  
A recent online report also indicates significant power increases for memory-intensive applications 
after an RDIMMs-to-MRDIMMs server upgrade~\cite{phoronix:MRDIMMs}. 
These observations have contributed to the prevailing perception 
that the performance gains enabled by MRDIMMs come at a significant power cost. 

Our detailed power characterization reveals a more complete picture and supports a different interpretation of these observations. 
Our analysis distinguishes between two bandwidth regions: the \textbf{shared window} (up to approximately 500\,GB/s), in which both RDIMM- and MRDIMM-based systems operate and therefore can be compared directly, and the \textbf{MRDIMM-extended window} (500--700\,GB/s), which is enabled by the RDIMM-to-MRDIMM upgrade. 

\begin{figure}[!t]
    \centering
    \includegraphics[width=0.99\columnwidth]{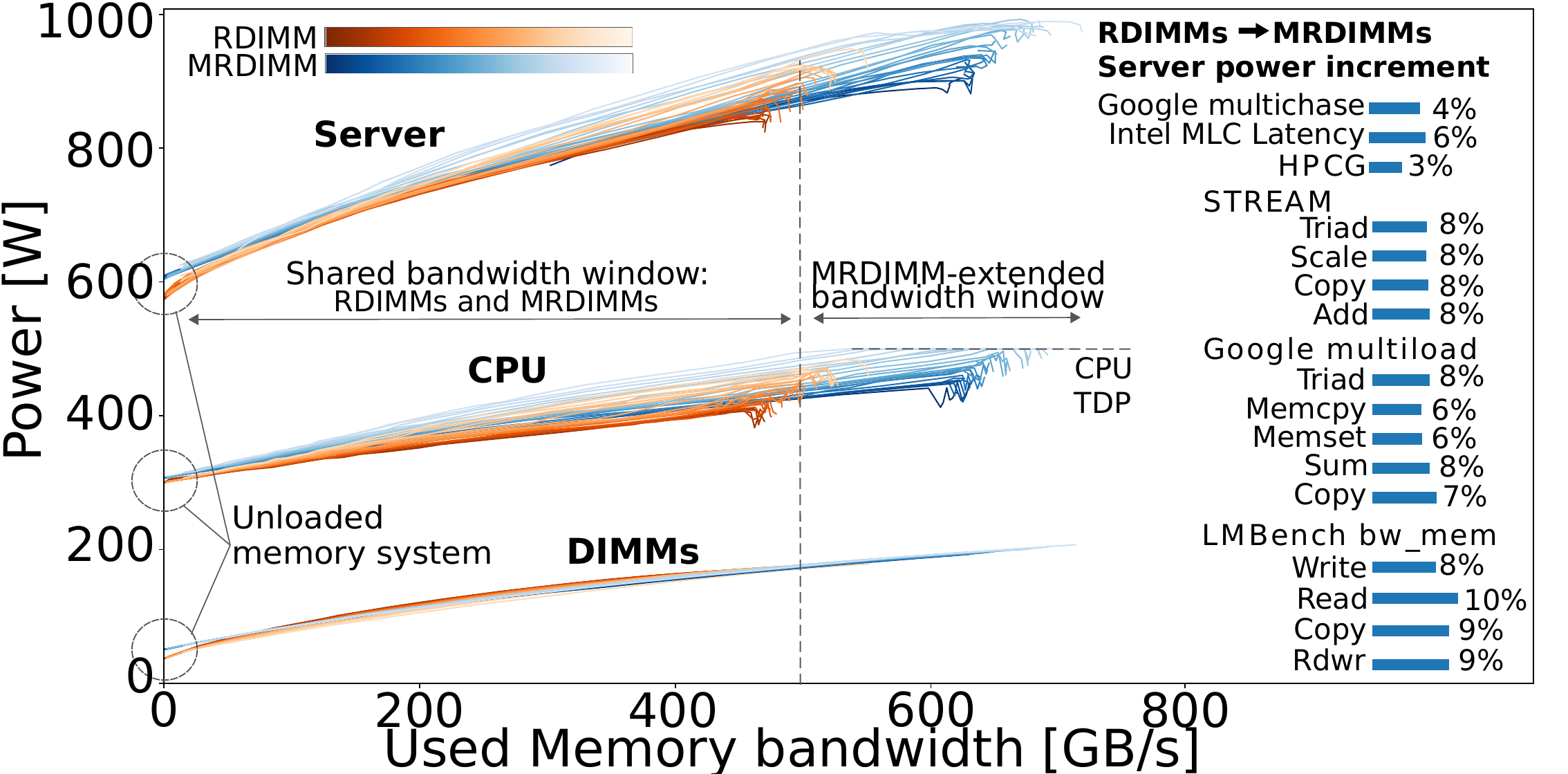}
    \caption{Across most of the shared bandwidth window, below 500\,GB/s, upgrading to MRDIMMs increases total server power by only 10--20\,W. In the MRDIMM-extended window, 500--700\,GB/s, the system sustains higher memory bandwidth, delivers higher overall performance, and as expected, consumes more power.}    
    \label{fig:mrdimms-vs-rdimms:power}
\end{figure}

%%%%%%%%%%%%%%%%%%%%%%%%%%%%%%%%%%%%%%%%% 
% Describe the power figure  
%%%%%%%%%%%%%%%%%%%%%%%%%%%%%%%%%%%%%%%%% 
Fig.\,\ref{fig:mrdimms-vs-rdimms:power} shows the bandwidth--power curves for the DIMMs, CPU and the 
entire server for the RDIMM-based~(orange) and MRDIMM-based~(blue) systems. 
The bar chart shows the server-power increase for the benchmarks under study.   
The bandwidth position of the benchmarks on the power curves, not shown explicitly in Fig.\,\ref{fig:mrdimms-vs-rdimms:power}, % closely 
corresponds to their position  
in Fig.\,\ref{fig:mrdimms-vs-rdimms:performance}. 
Google Multichase and Intel MLC latency are located in the unloaded-memory region, whereas the remaining benchmarks fully utilize the available memory bandwidth in both RDIMM- and MRDIMM-based systems. 

%%%%%%%%%%%%%%%%%%%%%%%%%%%%%%%%%%%%%%%%% 
% Shared RDIMM & MRDIMM bandwidth window 
%%%%%%%%%%%%%%%%%%%%%%%%%%%%%%%%%%%%%%%%% 

In the unloaded memory system, the power consumption of 12$\times$64\,GB MRDIMMs exceeds that of RDIMMs by approximately 15\,W,  leading to 4\% and 6\% higher power consumption 
of the Google Multichase and Intel MLC latency benchmarks. % (horizontal bar chart in Figure~\ref{fig:mrdimms-vs-rdimms:power}). 
This increase is likely attributable to the additional multiplexed data buffers, which are not present in RDIMMs, as well as to the greater complexity of the multiplexed RCD.

Across most of the shared bandwidth window, below 500\,GB/s, upgrading to MRDIMMs increases total server power by only 10--20\,W, an increment so small that it is hardly visible in the server bandwidth--power curves in Fig.\,\ref{fig:mrdimms-vs-rdimms:power}. This minor increase originates primarily from a slightly higher CPU power consumption (CPU bandwidth--power curves), most likely because of the higher memory-controller frequency. In contrast, MRDIMMs exhibit \textbf{the same or even lower power} consumption than RDIMMs at the same level of memory-bandwidth utilization: in the shared bandwidth window, the RDIMM and MRDIMM bandwidth--power curves practically overlap.   
As discussed in the previous paragraph, the MRDIMMs require some additional power because of multiplexed data buffers and RCDs. However, this power is compensated by DRAM chips operating \textbf{more efficiently} at a lower frequency of 4400\,MT/s in MRDIMM--8800 devices, compared to 6400\,MT/s in RDIMMs~\cite{rambus:mrdimms}.

%%%%%%%%%%%%%%%%%%%%%%%%%%%%%%%%%%%%%%%%% 
% MRDIMM-enabled bandwidth window: More perf, more power. 
%%%%%%%%%%%%%%%%%%%%%%%%%%%%%%%%%%%%%%%%% 
In the MRDIMM-extended bandwidth window, the server sustains higher memory bandwidth.  
As expected, this requires a higher power consumption, which is 
distributed across the DRAM DIMMs~(20--35\,W), CPU~(30--40\,W) and the remaining server components (below 20\,W). 
The power increment is also visible in the bandwidth-bound benchmarks,  
whose server power consumption increases between 3\%~(HPCG) and 10\%~(LMBench Read), 
as shown in the bar chart of Fig.\,\ref{fig:mrdimms-vs-rdimms:power}. 

Our detailed power evaluation reveals one unexpected finding:  
at the highest levels of memory bandwidth utilization, CPU power decreases by 20--35\,W. 
This reduction is visible in the rightmost segments of both the CPU and the total-system power curves in Fig.\,\ref{fig:mrdimms-vs-rdimms:power}.   
This is the first study to report this CPU-power behavior, whose detailed analysis is part of the ongoing work. 
As initial steps in this investigation, we identified two factors that are strongly correlated with this power reduction.
The first one is the activation of the write-allocate-evasion policy,  
which allows high-end Intel CPUs to bypass cache allocation on certain store misses despite the architectural model being write-allocate~\cite{ICX:SpecI2M}. 
The second is a steep increase in memory access latency, 
as observed in Fig.\,\ref{fig:mrdimms-vs-rdimms:performance}.

%% file: 50-Energy.tex
% -------------------------------------------------------------------------------
\section{Energy efficiency}
\label{sec:energy}
\begin{figure}[!t]
    \centering
    \includegraphics[width=0.99\columnwidth]{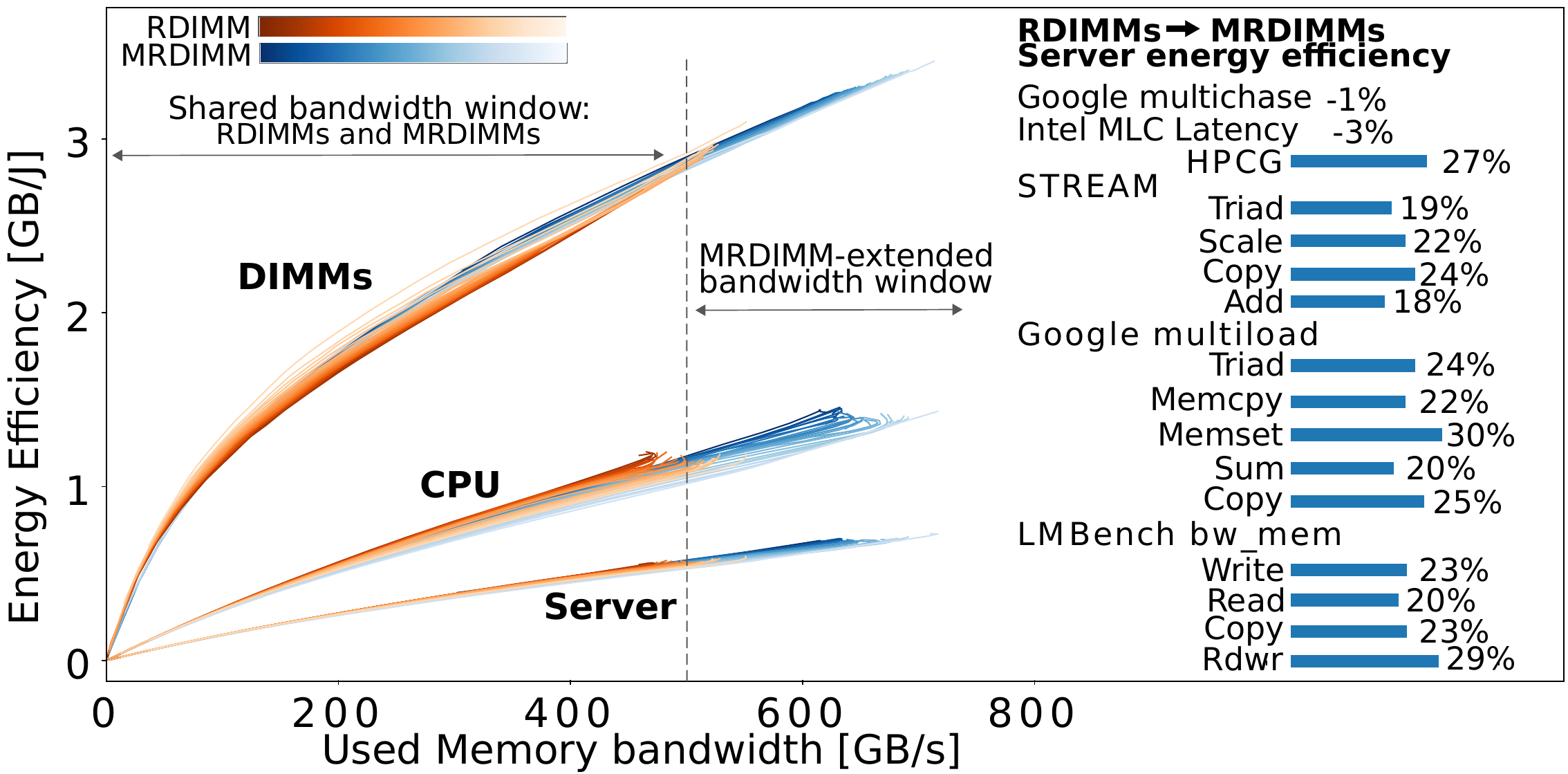}
    \caption{In the shared bandwidth window, the energy-efficiency curves of RDIMM- and MRDIMM-based systems largely overlap. The MRDIMM-extended window, 500--700\,GB/s, is the most energy-efficient operating region for the DRAM, CPU and the server as a whole. Bandwidth-bound benchmarks exploit this region, achieving 18--30\% higher energy efficiency.}
    \label{fig:mrdimms-vs-rdimms:energy}
\end{figure}

% -------------------------------------------------------------------------------
The detailed energy-efficiency characterization of RDIMM- and MRDIMM-based systems is shown in Fig.\,\ref{fig:mrdimms-vs-rdimms:energy}. 
The figure shows the energy efficiency (in GBs per Joule) for different memory-traffic intensity ($x$-axis) in 
RDIMM- and MRDIMM-based systems. The curves are based on the memory traffic generated by the Mess benchmarks, 
combined with the measurements of the DIMM, CPU and server energy counters.  
The horizontal bar chart shows the energy efficiency improvement 
of the memory-bound benchmarks because of the RDIMM-to-MRDIMM system upgrade. 

 In the unloaded-memory region (\tinytilde0\,GB/s), the server with MRDIMMs exhibits slightly lower energy efficiency, by less than 3\%. This difference is small enough to be barely visible in the energy-efficiency curves, but it can be observed in the 1--3\% lower efficiency of the Google multichase and Intel MLC Latency benchmarks. 
In the shared bandwidth window the energy-efficiency curves of RDIMM- and MRDIMM-based systems largely overlap. 
In the MRDIMM-extended window (500--700\,GB/s), energy-efficiency continues to increase with the used bandwidth, making it the most efficient operating region for the DRAM, CPU and the server as a whole.  
The bandwidth-bound benchmarks successfully exploit this operating region, resulting in their 18--30\% higher energy efficiency.

%% file: 60-MRDIMM-adoption.tex
% -------------------------------------------------------------------------------
\section{Cost and adoption}
\label{sec:MRDIMM-adoption}
% -------------------------------------------------------------------------------

Our results indicate that MRDIMMs are a promising technological innovation. Their extent and pace of adoption in production, however, will also depend on device cost and on our ability to identify application domains that can benefit substantially from this technology.

The current retail prices of the memory devices evaluated in our study, 64\,GB RDIMMs--6400 and MRDIMM--8800, are practically the same~\cite{mouser}. We also discussed the pricing of forthcoming Gen. 2 MRDIMM devices with a major memory manufacturer. 
Although no precise pricing data were available for these future products, 
the significant performance uplift and higher energy efficiency for bandwidth-bound workloads, 
as quantified by our study, are expected to be large enough to materially influence memory-selection decisions in modern servers.  
Even if next-generation MRDIMMs enter the market at a premium price, reflecting their more advanced buffering and signaling architecture, their added capability can still be economically justified, particularly in workloads where memory bandwidth or memory-access behavior is the dominant bottleneck. Beyond the memory-bandwidth gains, in environments such as large-scale AI inference, real-time analytics, fintech systems, and high-performance computing, higher per-node throughput often translates directly into server consolidation. Reducing the number of required servers lowers spending on compute, networking, software licensing, and data-center footprint. When these system-level savings are taken into account, MRDIMMs can enable a significantly lower total cost of ownership.

%% file: 70-Conclusions.tex
\section{Conclusions}
\label{sec:conclusions}

This paper presented a detailed performance, power and energy evaluation of a production server with high-end MRDIMM main memory. 
The memory system upgrade from conventional RDIMMs--6400 to MRDIMMs--8800 extends the bandwidth by 41\% 
yielding 27--41\% higher performance for bandwidth-bound workloads. 
Also, the latency improvements reach hundreds of nanoseconds, 
benefiting a broad class of workloads sensitive to memory latency. 
At the same bandwidth utilization levels, RDIMMs and MRDIMMs exhibit similar power consumption.   
In the MRDIMM-extended bandwidth region, the performance improvements largely exceed the power increase,  
delivering up to 30\% server energy savings for memory-bound workloads.
This performance and energy-efficiency uplift is large enough to materially influence memory-selection decisions in modern servers, even if these advanced memory devices enter the market at a premium price.

%% file: refs.bib
@misc{mouser,
Howpublished = {\url{https://www.mouser.com/}},
Title = {{Mouser Electronics}},
Year = {2026}}

@misc{Intel:MRDIMM-news,
Howpublished = {\url{https://newsroom.intel.com/data-center/new-ultrafast-memory-boosts-intel-data-center-chips}},
Title = {{New Ultrafast Memory Boosts Intel Data Center Chips}},
Month = {Nov}, 
Year = {2024}
}

@ARTICLE {Micron:MRDIMMs,
author = {{Micron Technology Inc.}},
journal = {Micron White paper},
title = {{Unlock the power of more cores with MRDIMM}},
year = {2024}
}

@ARTICLE {Lenovo:MRDIMMs,
author = {{Lenovo Press}},
title = {{Introduction to MRDIMM Memory Technology}},
year = {2025}
}

@misc{phoronix:MRDIMMs,
Howpublished = {\url{https://www.phoronix.com/review/ddr5-6400-mrdimm-8800}},
Title = {{Revisiting DDR5-6400 vs. MRDIMM-8800 Performance With Intel Xeon 6 ``Granite Rapids''}},
Month = {September}, 
Year = {2025}
}

@article{Dravai:MRDIMMs,
title = {{Performance and efficiency: A multi-generational benchmark of modern processors on bandwidth-bound HPC applications}},
journal = {Future Generation Computer Systems},
year = {2025},
author = {B. Drávai and I. Z. Reguly}
}

@ARTICLE{JEDEC:JESD82-521_2021,
  author = {{JEDEC}},
  title = {{JESD82-521: DDR5 Data Buffer Definition (DDR5DB01). Revision 1.1}},
  year = {December 2021}
}

@ARTICLE{JEDEC:JESD82-514_2024,
  author = {{JEDEC}},
  title = {{JESD82-514.01: DDR5 Registering Clock Driver Definition~(DDR5RCD04)}},
  year = {June 2024}
}

@article{rambus:mrdimms,
  title={{Expanding Server Memory Capabilities with Multiplexed Rank DIMM (MRDIMM) Technology}},
  author={Z. Mollah},
  journal={White paper},
  publisher = {Rambus inc.},
  year={2025}
}

@online{Intel:MLC,
  author = {{Intel Corporation}},
  title = "Intel {M}emory {L}atency {C}hecker v3.5",
  url  = "https://software.intel.com/en-us/articles/intelr-memory-latency-checker",
  Howpublished = {\url{https://software.intel.com/en-us/articles/intelr-memory-latency-checker}},
  year = 2023,
}

@article{mccalpin:streamBenchmark,
  title={{Memory bandwidth and machine balance in current high performance computers}},
  author={McCalpin, J. D.},
  journal={IEEE TCCA},
  year={1995}
}

@misc{lmbench,
	Howpublished = {\url{http://lmbench.sourceforge.net}},
	Title = {{LMbench}},
	Year = {2005}}

@misc{Google:multichase,
	title = "Multichase",
	author = "Google",
	howpublished = "\url{https://github.com/google/multichase}",
	year = "2021"
}

@online{xeon-6980p,
    title = {\textregistered{I}ntel  \textregistered{X}eon {6980P Processor}},
    author = {{I}ntel {C}orporation},
    year = {September 2024},
    url = {https://www.intel.com/content/www/us/en/products/sku/240777/intel-xeon-6980p-processor-504m-cache-2-00-ghz/specifications.html}
}

@inproceedings{esmailidokht2024mess,
  author    = {P. Esmaili-Dokht and others},
  title     = {{A Mess of Memory System Benchmarking, Simulation and Application Profiling}},
  booktitle = {MICRO},
  year      = {2024}
}

@inproceedings{rapl,
  author    = {H. David and others},
  title     = {{RAPL: Memory power estimation and capping}},
  booktitle = {ISLPED},
  year      = {2010}
}

@misc{ipmitool,
	title = "{IPMI}tool",
	author = "D. Laurie and others",
	howpublished = "\url{https://github.com/ipmitool/ipmitool}",
}

@ARTICLE{xeon_6,
  author={Powell, M. D. and others},
  journal={IEEE Micro}, 
  title={{Intel Xeon 6 Product Family}}, 
  year={2025}}

@INPROCEEDINGS{ICX:SpecI2M,
  author={Papazian, I. E.},
  booktitle={IEEE Hot Chips 32 Symposium}, 
  title={New 3rd {G}en {I}ntel® {X}eon® {S}calable {P}rocessor ({C}odename: {I}ce {L}ake-{SP})}, 
  year={2020},
  doi={10.1109/HCS49909.2020.9220434}}

@techreport{intel:hpcg,
  author    = {{Intel Corporation}},
  title     = {{oneMKL Developer Guide for Linux}},
  pages     = {175},
}
